\documentclass[onecolumn,a4paper,draftclsnofoot]{IEEEtran}

\usepackage{cite}	% citation
\usepackage{graphicx} % grafics, floats (figs)
\DeclareGraphicsExtensions{.pdf,.eps}
\usepackage[latin1]{inputenc} % language
\usepackage[T1]{fontenc} % language and fonts
\usepackage{amsmath,amsfonts,amsbsy,amssymb} % mathematical lybraries
\usepackage{mathabx} 
\usepackage{mathtools}
\usepackage{amssymb} 
\usepackage{amsmath}
\usepackage{amsthm}	
\usepackage{mathrsfs}
\usepackage[nolist]{acronym} % acronym lybrary
\usepackage{tabularx} % tables
\usepackage{multirow}
\usepackage{wasysym}
\usepackage{float}
\usepackage{color} % colors (fonts and edting)

\usepackage{enumitem} % special package for enumeration

\begin{document}
	
	\title{Ultra Reliable Communication via Optimum Power Allocation for Type-I ARQ in Finite Block-Length}
	\author{Endrit Dosti, Uditha Lakmal Wijewardhana, Hirley Alves and Matti Latva-aho\\
		\IEEEauthorblockA{
			Centre for Wireless Communications (CWC), University of Oulu, 90014 Oulu, Finland\\
		}
				
		email:firstname.lastname@oulu.fi  
	}
	%
	% - - - - - - - - - - - - - - - - - - - - - - - - - - - - - - - - - - - - - - - - - -
	\maketitle
	
	% - - - - - - - - - - - - - - - - - - - - - - - - - - - - - - - - - - - - - - - - - -
	
\author{
	\IEEEauthorblockN{, , , and } 

}
%
% - - - - - - - - - - - - - - - - - - - - - - - - - - - - - - - - - - - - - - - - - -
\maketitle
	\begin{abstract}
		
		We analyze the performance of the type-I automatic repeat request (ARQ) protocol with ultra-reliability constraints. First, we show that achieving a very low packet outage probability by using an open loop setup is a difficult task. Thus, we introduce the ARQ protocol as a solution for achieving the required low outage probabilities for ultra reliable communication. For this protocol, we present an optimal power allocation scheme that would allow us to reach any outage probability target in the finite block-length regime. We formulate the power allocation problem as minimization of the average transmitted power under a given outage probability and maximum transmit power constraint. By utilizing the Karush-Kuhn-Tucker (KKT) conditions, we solve the optimal power allocation problem and provide a closed form solution. Next, we analyze the effect of implementing the ARQ protocol on the throughput. We show that by using the proposed power allocation scheme we can minimize the loss of throughput that is caused from the retransmissions. Furthermore, we analyze the effect of the feedback delay length in our scheme.

	\end{abstract}
	
	\section{Introduction}
	\label{Introduction}
	Mobile communication systems are becoming more and more important in everyone's daily life. So far the main focus of these systems has been to provide higher data rates to the end users. This trend needs to be continued in the next generation of mobile communication systems, namely 5G, while introducing new features for a better user experience \cite{Popovski2014}. One of these features is the Internet of Things (IoT), which has gathered much attention from both research community and industry. The IoT interconnects "things" (e.g machines, smart meters and everyday objects) and autonomously exchanges data between them\cite{Networks}. This may enable the creation of many services and business models.  
	
	Machine type communication (MTC)  will pave path for seamlessly and ubiquitous connectivity foreseen in 5G and IoT \cite{Palattella,Andreev}. The MTC networks have a massive number of devices communicating with diverse range of requirements in terms of reliability, latency, data rates, and energy consumption, besides diverse traffic patterns \cite{Nokia}. Thus, in this context, these objects (e.g. machines) will be interconnected with different requirements. In this paper we focus on those devices that must be connected to the wireless network almost all the time. Therefore, we require the communication link to have a probability of successful transmission around 99.999\% or higher. This operation mode is known as ultra reliable communication. Most of these devices in a MTC network are expected to be rather simple, such as sensors, and the amount of information they transmit will be small. Despite their size, the information that is transmitted from these devices needs to be decoded at the intended receiver with very high reliability and very low latency (which depend on the system specifications). 
	
One of the major challenges in wireless communications is the presence of fading, which causes fluctuations in the received signal power, thus resulting in loss of transmitted packets \cite{Goldsmith}. Automatic repeat request (ARQ) protocols  can be used to mitigate this effect. Therein, we transmit the same packet in several independent fading paths and then combine them accordingly at the receiver. Moreover, ARQ protocols are well-known in achieving the desired reliability level while maintaining certain complexity \cite{Larsson}. Through ARQ, the outage probability of a wireless communication link is reduced by retransmitting the packets that experienced bad channel conditions.
	
	The problem of implementing the ARQ protocol or other retransmission schemes together with power allocation schemes has been investigated in \cite{Larsson, Chaitanya2016, Tumula2}, but it is done under the assumption of asymptotically long codewords. This implies that the length of metadata (control information) is much smaller than the actual data, thus the performance metrics that are used are the channel capacity and its extension to non-ergodic channels, the outage capacity. In the finite block-length regime metadata and the actual data are almost of the same size, therefore these conventional methods are highly suboptimal \cite{Durisi_1}. Thus, the performance of these schemes with power allocation should be evaluated. Little work has been done in this field for the short packets domain. For instance, \cite{Kim2013} evaluates the performance of incremental redundancy ARQ for additive white Gaussian noise (AWGN) channel  showing that a large number of retransmissions enhances performance in terms of the long term average transmission rate (LATR). This scheme requires a more sophisticated encoding/decoding mechanism that may not be feasible in nodes with limited computational capabilities. Furthermore,  authors do not assess the impact of such large number of retransmissions on latency nor do they guarantee high reliability. Further, in \cite{Devassy2014} the authors analyze the performance of ARQ protocol over the fading channel under the assumption of infinite number of transmissions and an instantaneous and error free feedback. Moreover, the impact of power allocation between different ARQ rounds has not been analyzed. In \cite{Makki2015}, the authors develop a power allocation scheme that minimizes the outage probability but only for the case of two retransmissions. However, in their algorithm they do not guarantee  a minimal outage probability level which would be essential in the case of ultra reliable communications, since different applications have different reliability requirements.
	
	In this paper, we develop a power allocation scheme for type-I ARQ protocol that allows us to achieve a target reliability level under fading channels. We analyze the performance for the general case of $M$ transmissions. Then we investigate the impact of implementing this ARQ scheme on the throughput of the wireless network. We show that the optimal power allocation scheme allows us to minimize the losses in throughput that are caused by the presence of retransmissions. Finally, we evaluate how throughput behaves as a function of ARQ feedback delay and show that the loss is relatively low for a wide range of delays.

	\section{System model} 
	\label{sc:system_model}
	
	Assume a transmitter-receiver pair communicating under an ARQ protocol, where the maximum number of ARQ rounds is set	to $M$, where in each round we will transmit a scaled version	of the initial codeword. Furthermore we assume that the receiver does not buffer all the received packets. This implies that if it can not decode a certain packet, it drops that packet and requests for retransmission. If at the end of the $M$-th round, transmission still is not successful, the packet is dropped and error is declared.   
 	
 	We consider quasi-static fading channel conditions, in which the channel gain $h$ remains constant for the duration of one packet transmission and changes independently between ARQ rounds. We analyze the case when the channel coefficient is Rayleigh distributed and $h\sim \mathcal{CN} (0,1)$. Thus the squared envelope of the channel gain is exponentially distributed with mean one. For simplicity we denote  $f_{|h|^2} (z) = \exp(-z)$. We assume that the receiver has channel state information while the transmitter knows only the distribution of the channel gains and the information it obtains from the feedback. Then the received signal at the $m^{th}$ round can be written as
	\begin{align}
	\label{eq:1}
	y_m = \sqrt{\rho_m}h_mx_{m} + w_{m},
	\end{align}
	where $x_{m}$ is the transmitted signal and $w_{m}$ is the AWGN noise term with noise power $N_0=1$. The term $\rho_m$ is the packet transmit power, which since the variance of the noise is set to 1, corresponds to the transmission signal-to-noise ratio (SNR).

	\section{Maximum coding rate in finite block-length}
	\label{3}
	In this section, we briefly summarize the recent results in the characterization of the maximum channel coding rate and outage probability in the finite block-length regime. Further, we evaluate the case of the open loop setup. 
	
	For notational convenience we need to define an $(n, K, \rho, \epsilon)$ code as a collection of
	\begin{itemize}
		\item An encoder $\mathcal{F} : \{1, \ldots, K\} \mapsto \mathcal{C}^n $ which maps the message $ k \in \{1, \ldots, K\}$  into an $n$-length codeword $c_i \in \{c_1, \ldots , c_n\}$ such that the following power constraint ($\rho$) is satisfied:
			\begin{align}
		\label{eq:2}
	\frac{1}{n} \|c_i\|^2 \leq \rho, \forall i.
		\end{align}
		\item A decoder $\mathcal{G} : \mathcal{C}^n \mapsto \{1, \ldots, K\}$ that satisfies the maximum error probability ($\epsilon$) constraint:
		\begin{align}
		\label{eq:3}
		\max_{\forall i} \mathrm{Pr}\left[\mathcal{G}(y) \neq I|I=i\right] \leq \epsilon,
		\end{align}
		where $y$ is the channel output induced by the transmitted codeword according to (\ref{eq:1}).
	\end{itemize}
	The maximum achievable rate of the code is defined as:
	\begin{align}
	\label{eq:4}
	R_{max}^*(n, \rho, \epsilon)=\sup \Big\{\frac{\log  K}{n}: \exists(n, K, \rho, \epsilon)~\mathrm{code}\Big\},
	\end{align}
	where $\log$ refers to the natural logarithm. 
	
	For the AWGN channel non-asymptotic lower and upper bounds on the maximum achievable rate have been derived in \cite{Polyanskiy2010}. Recently, a tight approximation for $R_{max}^*(n, \rho, \epsilon)$ has been proposed for sufficiently large values of $n$ in the case of the quasi-static fading channel \cite{Yanga} and is given by 
	\begin{align}
	\label{eq:5}
	R_{max}^*(n, \rho, \epsilon)\approx C_\epsilon + \operatorname{O}\left( \frac{\log n}{n}  \right),
	\end{align}
	where $C_\epsilon$ is the outage capacity:
	\begin{align}
	\label{eq:6}
	C_\epsilon= \sup \{R: \mathrm{Pr}[\log (1+\rho \cdot z)<R]<\epsilon\}. 
	\end{align}
	Then, by assuming a code rate of $R=\frac{K}{n}$ nats per channel use (ncpu), where $K$ is the information payload, the outage probability is approximated as \cite{Yangk}
	\begin{align}
	\label{eq:7}
	\epsilon (n, R, \rho) &\approx E\left[Q\left(\frac{C(\rho \cdot z)-\frac{K}{n}}{\sqrt{V(\rho \cdot z)}}\right)\right] \\ &\approx \int_{0}^{\infty} e^{- z  } Q\left(\frac{C(\rho \cdot z  )-\frac{K}{n}}{\sqrt{V(\rho \cdot z  )}}\right)\mathrm{d}z,
	\end{align}
	where $E[\cdot]$ denotes the expectation over the channel gain $z $, $Q(\cdot)$ denotes the Gaussian Q-function, $C(x) = \log(1+x)$ denotes the channel capacity and the channel dispersion is computed as $V(x) = 1 - \tfrac{1}{(1+x)^2}$. However the integral in (8) does not have a closed form solution. Thus, we resort to an approximated closed-form expression as in \cite{Makki2014}
	 \begin{align}
	 \label{eq:11}
	 \epsilon (n, R, \rho) = 1-\frac{\delta}{\sqrt{2\pi}}e^{-\kappa}\left(e^{\sqrt{\frac{\pi}{2\delta^2}}}-e^{-\sqrt{\frac{\pi}{2\delta^2}}}\right),
	 \end{align}
	 where $\kappa=\frac{e^R-1}{\rho}$ and $\delta=\sqrt{\frac{n\rho^2}{e^{2R}-1}}$. Note that \eqref{eq:11} characterizes the outage probability of a single ARQ round.
	 
\begin{figure}[!htb] % [!t] or [!b] or [!h] % force fitting, force top, force bottom, force text fitting
		\centering
		\includegraphics[trim=0.8cm 0.1cm 1.3cm 0.7cm, clip=true,width=.8\linewidth]{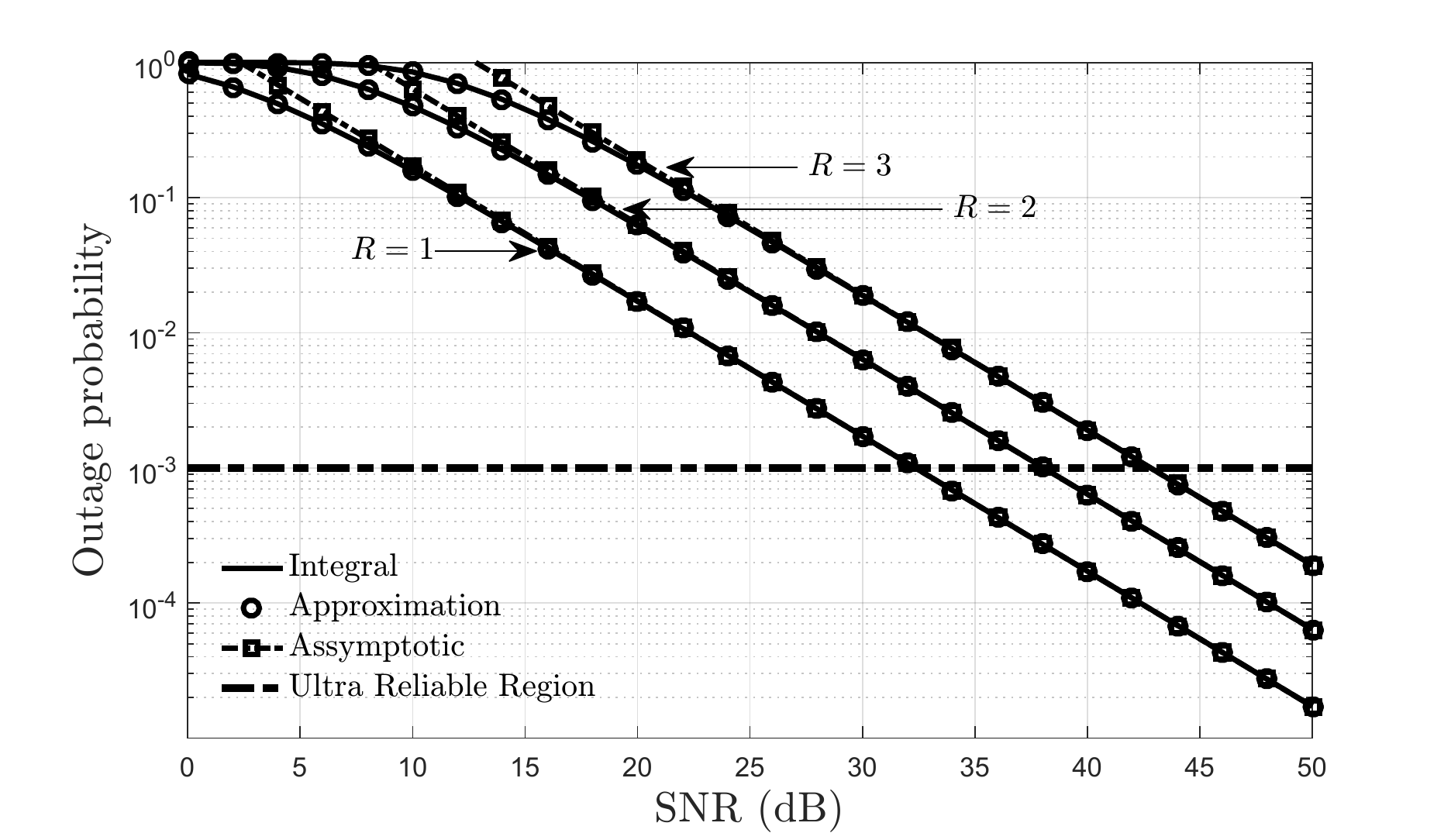}
		\caption{Outage probability for the open loop setup for different channel coding rates.}
		\label{fig:open_loop_setup}
		\vspace*{-3mm}
	\end{figure}

	 Fig. \ref{fig:open_loop_setup} illustrates the outage probability for the open loop setup, where the message is conveyed in a single transmission, for different channel coding rates. We have fixed the number of channel uses $n=200$ and analyzed the case of mapping $K \in \{200,400,600\}$ information nats. This results in the channel coding rates $R=1$, $R=2$ and $R=3$ ncpu, respectively. We can see that the integral form in (8) and the closed-form approximation in \eqref{eq:11} match well for all the coding rates.

	 Furthermore, in Fig. \ref{fig:open_loop_setup} we also illustrate the performance of the asymptotic approximation (which we derive next and is given by \eqref{eq:14}). The results show that in the  ultra reliable regime, where very low outages are required $\epsilon < 10^{-3}$, the asymptotic approximation can be used. This is due to the fact that in high SNR the maximum achievable rate (\ref{eq:5}) converges to the one with asymptotically long codewords $R^* (n,\epsilon)=C_{\epsilon}$, where $C_{\epsilon} $ is defined in (\ref{eq:6}). Therefore, when $\rho \rightarrow \infty$ the outage probability in the $m^{th}$ round can be calculated as:
	 \begin{align}
	 \label{eq:13}
	 \epsilon_m=\mathrm{Pr}[\log (1+\rho \cdot z  )<R]=1-e^{\frac{e^{R}-1}{\rho_m}}.
	 \end{align}
	 The equality in (\ref{eq:13}) holds for Rayleigh fading channels. Furthermore, by using the first order of Taylor expansion $e^{-x} \approx 1-x$ we can express the asymptotic approximation for the outage probability of the $m^{th}$ round as:
	 \begin{align}
	 \label{eq:14}
	 \epsilon_m=\frac{\phi}{\rho_m},
	 \end{align}
	where $\phi=e^{R}-1$. 
	 
	 	For ultra reliable communications, we require to have an outage probability $\epsilon$ very low, while spending as little power as possible. However, Fig. \ref{fig:open_loop_setup} shows that such low outage values are highly unlikely to be obtained using an open loop setup. Thus, we investigate the possibility of utilizing a retransmission mechanism, specifically the ARQ protocol with optimal power allocation,  in order to obtain an outage probability in the ultra reliable region. 
	 
	\section {Power allocation and throughput analysis}
	\label{4}
	In this section we evaluate the impact of the ARQ protocol on the outage probability of the user's transmission and the throughput. First, we propose an optimal power allocation scheme for the ARQ protocol that allows us to reach any target outage probability assuming when we can have up to $M$-transmissions. Then, we provide the expressions for analyzing the throughput of the system in case of ARQ with $M$-transmissions.
	
		\subsection{Optimal power allocation}
	
	The problem of interest is to achieve a target outage probability while spending as little power as possible for sending the information from the transmitter to the receiver. Since, we have multiple transmissions when using the ARQ protocol, one approach would be to allocate equal power in each round. This implies that given a certain power budget $\rho_{budget}$, the transmit power in the $m^{th}$ ARQ round would be $\rho_m=\frac{\rho_{budget}}{M}$. However, such simplistic approach is shown to be highly inefficient for very low outage probability values \cite{Makki2015}. Thus, we propose a power allocation algorithm in order to minimize the average transmit power of the transmitter which allocates different power levels in each ARQ-round. Bearing this in mind, the average transmitted power can be defined as
	\begin{align}
	\label{eq:15}
	\rho_{avg}=\frac{1}{M}\sum_{m=1}^M \rho_m E_{m-1}\mathrm{,}
	\end{align}
	where $M$ is the maximum number of ARQ rounds, $\rho_m$ is the power transmitted in the $m^{th}$ round and $E_{m-1}$ is the outage probability up to the $m-1$ round. Next, we calculate the probability that the packet can not be decoded correctly even after the maximum allowed number of retransmissions. We refer to this as packet drop probability (pdp), and it corresponds to the outage probability up to the $M^{th}$ round ($E_M$). Since we have assumed that all the transmissions of the packets experience independent fading conditions we can express $E_m$ as
	\begin{align}
	\label{eq:16}
	E_M=\prod_{m=1}^M \epsilon_m \mathrm{,}
	\end{align}
	where $\epsilon_m$ is the outage probability of the $m^{th}$ ARQ round and can be computed by \eqref{eq:11}, or asymptotically via \eqref{eq:14}. The outage probability before the first transmission, $\epsilon_0=1$. 
		
	Now, we can formulate the following power allocation problem:
	\begin{equation}
	\begin{aligned}
	\label {eq:17}
	& {\text{minimize}}
	& & \mathrm{\rho_{avg}} \\
	& \text{subject to}
	& & 0 \leq \rho_{m}, \; 1 \leq m \leq M \\
	&&& E_M = \epsilon
	\end{aligned}
	\end{equation}
	where $\epsilon$ is any target outage probability. Since the optimization problem (\ref{eq:17}) is non-linear and the feasible set is compact, we can find a global optimal solution \cite{5754756}. Here, we utilize the Karush-Kuhn-Tucker (KKT) conditions to obtain the optimal solution for the convex problem (\ref{eq:17}).
	
	We start by writing the Lagrangian function for problem \eqref{eq:17}, which is given by
	\begin{align}
	\label{eq:18}
	\mathcal {L} (\rho_m, \mu_m,\lambda) &= \frac{1}{M}\sum_{m=1}^M \rho_m E_{m-1} + \sum_{m=1}^M \mu_m \rho_m+ \lambda(E_M-\epsilon)
	\end{align}
	where $\mu_m$ for $m=1,\ldots,M$  and $\lambda$ are the Lagrangian multipliers. Now, we express the KKT conditions as follows:
		\begin{enumerate}[label=\textbf{C\theenumi},itemsep=2pt,parsep=2pt,topsep=2pt,partopsep=2pt]
		\item $\frac{\partial{\mathcal{L}}}{\partial {\rho_m}}=0,~ m=1,\ldots,M$,
		\label{c1}	
		\item $\mu_m \geq 0,~ m=1,\ldots,M$,
		\label{c2}
		\item $\mu_m\rho_m=0,~ m=1,\ldots,M$,
		\label{c3}
		\item $E_M-\epsilon=0$.
		\label{c4}
	\end{enumerate}
	
	We can write the derivative of the Lagrangian function $\mathcal {L} (\rho_m, \mu_m,\lambda)$ with respect to the power $\rho_m$ as
	\begin{align}
	\frac{\partial \mathcal {L} (\rho_m, \mu_m,\lambda)}{\partial \rho_m }=&\frac{1}{M}\left(\frac{\phi^{m-1}}{\prod_{i=1}^{m-1}\rho_i} - \sum_{i=1}^{M-m}\frac{\rho_{m+i}\phi^{m+i-1}}{\rho_m^2 \prod_{j=1, j \neq m}^{m+i-1}\rho_j}\right) %\nonumber\\&  
	- \mu_m - \lambda\left(\frac{\phi^M}{\rho_m^2 \prod_{i=1, i \neq m}^{M}\rho_i}\right).
	\label{eq:19}
	\end{align}
	The transmit power at $m^{th}$ ARQ round $\rho_m > 0$ for $m =1, \ldots,M$ since we require some power to transmit the information at each of these rounds. Thus, from the complementary slackness condition (\ref{c3}) we can see that $\mu_m = 0$ for $m =1,...,M$. Now, using \eqref{eq:19} and $\mu_M = 0$, we can write \ref{c3} for $m = M$ as 
	\begin{align}
	\label{eq:20}
	\frac{\partial \mathcal {L} (\rho_m, \mu_m,\lambda)}{\partial \rho_M }\!=\!\frac{1}{M}\frac{\phi^{M-1}}{\prod_{i=1}^{M-1}\rho_i} \!-\! \lambda\left(\frac{\phi^M}{\rho_M^2\prod_{i=1,i\neq m}^M \rho_m}\right)\!\!=\!0.
	\end{align}
	After some algebraic manipulations of \eqref{eq:20} we obtain the transmit power at the $M^{th}$ ARQ round as 
	\begin{align}
	\label{eq:21}
	\rho_M=\sqrt{ M\lambda \phi}.
	\end{align}
	Similarly, substituting $m=M-1$ in \eqref{eq:19} and using $\mu_{M-1} = 0$, we can rewrite \ref{c1} for $m= M-1$ as
	\begin{align}
	\label{eq:23}
	\frac{\partial \mathcal {L} (\rho_m, \mu_m,\lambda)}{\partial \rho_{M-1} }&=\frac{1}{M}\left(\frac{\phi^{M-2}}{\prod_{i=1}^{M-2}\rho_i}-\frac{\rho_M\phi^{M-1}}{\rho_{M-1}^2\prod_{i=1}^{M-2}\rho_i}\right) %\nonumber\\&- 
	-\lambda\left(\frac{\phi^M}{\rho_{M-1}^2\prod_{i=1,i\neq M-1}^M \rho_i}\right)=0.
	\end{align}
	Mathematical simplification of (19) leads to
	\begin{align}
	\label{eq:41}
	\rho_{M-1}=\sqrt{\rho_M\phi+\frac{M\lambda\phi^2}{\rho_M}}.
	\end{align}
	Following a similar approach, we obtain the following relationship for the case of $m = M-2$ 	
	\begin{align}
	\label{eq:43}
	\rho_{M-2}=\sqrt{\rho_{M-1}\phi+\frac{\rho_M\phi^2}{\rho_{M-1}}+\frac{M\lambda\phi^3}{\rho_{M-1}\rho_M}}.
	\end{align}
	We can continue this procedure for all $m \in \{1, \ldots, M\}$ and the results can be summarized as:
	\begin{align}
	\rho_M&=f(\lambda)\\
	\label{b}
	\rho_{M-1}&=f(\lambda, \rho_M) \\
	\label{a}
	\rho_{M-2}&=f(\lambda, \rho_M, \rho_{M-1}) \\
	\vdotswithin{rho_{1}} \nonumber\\
	\rho_{1}&=f(\lambda, \rho_M, \ldots, \rho_3, \rho_2) 
	\end{align}
	 
	 By utilizing a method that is
	 similar to the backward substitution approach \cite[App. C.2]{Boyd-Vandenberghe-04}, we can obtain a relationship between the power terms $\rho_m$ as follows: first, by substituting $M \lambda \phi = \rho_M^2$ (see (\ref{eq:21})) in \eqref{eq:41} (or equivalently in \eqref{b}) we evaluate $\rho_{M-1}$ as $\rho_{M-1} = \sqrt{2\phi \rho_M}$. Next, $\rho_{M-2}$ is evaluated by substituting  $\sqrt{M\lambda \phi} = \rho_M$ and  $\sqrt{2\phi \rho_M} = \rho_{M-1}$ in \eqref{a}. By continuing this procedure we can express the optimal transmit power in the $m^{th}$ round as
	\begin{align}
	\label{eq:44}
	\rho_m=\sqrt{2\phi\rho_{m+1}}, \; 1\leq m<M.
	\end{align}
	Since $\rho_M$ is a function of $\lambda$ (see \eqref{eq:21}) and using \eqref{eq:44}, it is clear that each $\rho_m$ is a function of $\lambda$. Thus, all that remains is to compute the Lagrangian multiplier $\lambda$. For this purpose, we utilize the outage constraint in (\ref{eq:17}) (\ref{c4}). First, we substitute $\rho_m$ for $m =1,\ldots,M$ in \eqref{eq:16} to obtain $E_M$ as
	\begin{align}
	\label{eq:90}
	E_m=\frac{\phi^M}{\prod_{m=1}^M \rho_m},
	\end{align}
	where $\rho_m$ is given by 
	\begin{align}
	\label{eq:91}
	\rho_m=\sqrt{2^a  \phi^b (M \lambda)^c}.
	\end{align}
	In \eqref{eq:91}, $a={2-2^{-(M-m-1)}}$, $b={2-2^{-(M-m)}}$ and $c={{2^{-(M-m)}}}$. Then, we solve for $\lambda$ by equating $E_M$ to the outage target $\epsilon$ as in \ref{c4}.

	\subsection{Throughput analysis}
	Utilizing schemes that are based on retransmissions causes the latency of a system to increase due to the presence of feedback. Furthermore, type-I ARQ schemes are also associated with a degradation of the spectral efficiency. These are then reflected in the overall system throughput. In this section, we analyze the impact of type-I ARQ scheme on the throughput for the general case of $M$ retransmissions. 
	
	In the open loop setup, which was analyzed in Section \ref{3}, the throughput ($\eta$) can be computed as
	\begin{align}
	\label{eq:26}
	\eta=\frac{k}{n}\left(1-\epsilon (n, R, \rho) \right).
	\end{align}
	However, during our analysis we need to consider the fact that after each ARQ round there exists a degradation of spectral efficiency. As a result of this degradation the spectral efficiency up to the $M^{th}$ ARQ round is given by $R_M=\frac{K}{Mn}$.  We evaluate the total number of channel uses until the $M^{th}$ round as
	\begin{align}
	\label{eq:27}
	\mathcal {T}=\sum_{m=1}^{M} mnE_{m-1}+D\sum_{m=1}^{M-1} mE_{m-1},
	\end{align}
	where $E_m$ is the packet drop probability given in (\ref{eq:16}) and $E_0=1$. Moreover, $D$ is the feedback delay expressed in channel uses. Note that $D$ accounts for the feedback delay from the NACK transmission, which are also assumed to use a few hundred channel uses. As pointed out in \cite{Durisi_1}, this models more accurately the impact of the retransmission protocols in practical scenarios compared to the conventional one bit feedback. Further, we can compute the expected number of information nats that will be transmitted as
	\begin{align}
	\label{eq:30}
	\mathcal {K}=K(1-E_M)=K\left(1-\prod_{m=1}^M \epsilon_m\right).
	\end{align}
	
	Thus, the throughput for the case of $M$ transmissions can be expressed as:
	\begin{align}
	\label{eq:31}
	\eta=\frac{\mathcal {K}}{\mathcal {T}}=\frac{K\left(1-\prod_{m=1}^M \epsilon_m\right)}{\sum_{m=1}^{M}mn E_{m-1}+D\sum_{m=1}^{M-1} mE_{m-1}}.
	\end{align}
	It is obvious from \eqref{eq:31} that the presence of feedback causes loss in the throughput. This degradation becomes worse as we increase the number of retransmissions. However, the proposed power allocation scheme helps to mitigate this effect as shown in the next section.
	
	\section{Numerical analysis}	\label{6}
	From Section \ref{4}.A it is clear that the analytical solution to obtain the optimal power allocation scheme for a large number of retransmissions would become a cumbersome task. Furthermore, in general the ultra reliable systems are delay-limited. Hence, in this section we limit our analytical analysis to the case of having a maximum of $M=2$ transmissions. Moreover, we evaluate numerically the impact of $M=3$ and $M=4$ transmissions in the system throughput. 

For the scenario of $M=2$ transmissions the optimization problem \eqref{eq:17} simplifies to
	\begin{equation}
	\begin{aligned}
	\label {eq:42}
	& {\text{minimize}}
	& & \mathrm{\frac{1}{2}(\rho_1+\rho_2\epsilon_1)} \\
	& \text{subject to}
	&& \frac{\phi^2}{\rho_1\rho_2} = \epsilon
	\end{aligned}
	\end{equation}
	
	{\noindent}To solve problem \eqref{eq:42} we can utilize the procedure described in Section \ref{4}.A. First, by using \eqref{eq:21} and \eqref{eq:44}  we compute the power terms as functions of $\lambda$. Next, by substituting these expressions for $\rho_1$ and $\rho_2$ in \ref{c4} and solving for $\lambda$ we obtain $\lambda=\frac{\phi}{2\epsilon}\sqrt[3]{\tfrac{1}{4\epsilon}}$. Finally, we compute the values of the power terms as $\rho_1=\phi\sqrt[3]{\tfrac{2}{\epsilon}}$ and $\rho_2=\frac{\phi}{\epsilon}\sqrt[3]{\tfrac{\epsilon}{2}}$. For this specific case of $M=2$ we can also utilize the following simpler approach to find the optimal power allocation. First, we rewrite the equality constraint as $\rho_2=\frac{\phi^2}{\rho_1 \epsilon}$. Next, by substituting $\rho_2$ in the objective function of \eqref{eq:42}, we obtain an unconstrained optimization problem with variable $\rho_1$. Then, we compute $\rho_1$ by setting the first derivative of the new objective function to zero as
	\begin{align}
	\label{eq:100}
	\frac{1}{2}-\frac{\phi^3}{\rho_1^3 \epsilon}=0.
	\end{align}
	By solving \eqref{eq:100} we find $\rho_1=\phi\sqrt[3]{\tfrac{2}{\epsilon}}$, which is same as what we obtained by using the procedure described in Section \ref{4}.A. Then, after substituting the first power term equation in the rewritten equality constraint we compute $\rho_2=\frac{\phi}{\epsilon}\sqrt[3]{\tfrac{\epsilon}{2}}$.
	
	\begin{figure}[!t] % [!t] or [!b] or [!h] % force fitting, force top, force bottom, force text fitting
		\vspace*{-3.5mm}
		\centering
		\includegraphics[trim=0.8cm 0.1cm 1.3cm 0.7cm, clip=true,width=.8\linewidth]{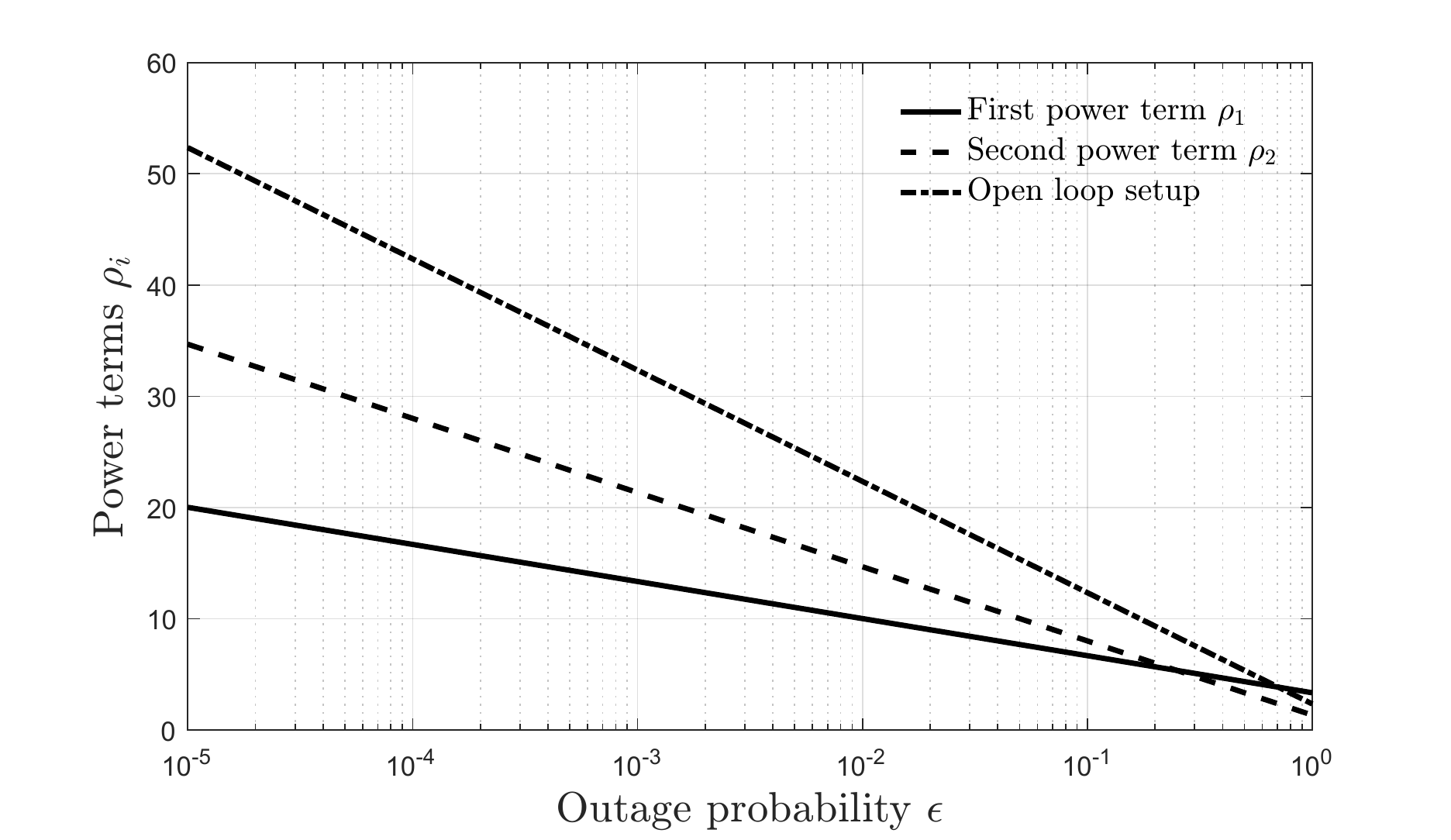}
		\caption{Transmit power in each ARQ round to achieve a target outage probability for rate $R=1$ ncpu.}
		\label{fig:Power_terms_vs_outage}
		\vspace*{-4mm}
	\end{figure}
	
	In Fig. \ref{fig:Power_terms_vs_outage} we illustrate the variation of transmit power $\rho_m$ in each ARQ round versus the outage target $\epsilon$. The results show that both power terms are lower than the open loop transmission which is shown in Fig.\ref{fig:open_loop_setup}. Furthermore, notice that if the first ARQ round is successful (i.e when the channel conditions are good), then the power gain with respect	to the open loop setup would be over 20 dB for $\epsilon \leq 10^{-3}$, which corresponds to the ultra reliable region. We observe that in this region, the first power term is lower than the second power term. This holds even for the general case of M transmissions and the mathematical proof will be shown later in a journal version. Notice that this result fully matches the intuition. Since our goal is to achieve a target outage probability by spending as less power as possible and there is no delay limitation, we transmit first with low power. If the channel conditions are good then the transmission will be successful, and we save a large amount of power. If we fail, then we retry until succeed or the maximum allowed number of retransmissions is reached.
	
\begin{figure}[!h]% [!t] or [!b] or [!h] % force fitting, force top, force bottom, force text fitting
		\centering
		\includegraphics[trim=0.8cm 0.1cm 1.3cm 0.7cm, clip=true,width=.8\linewidth]{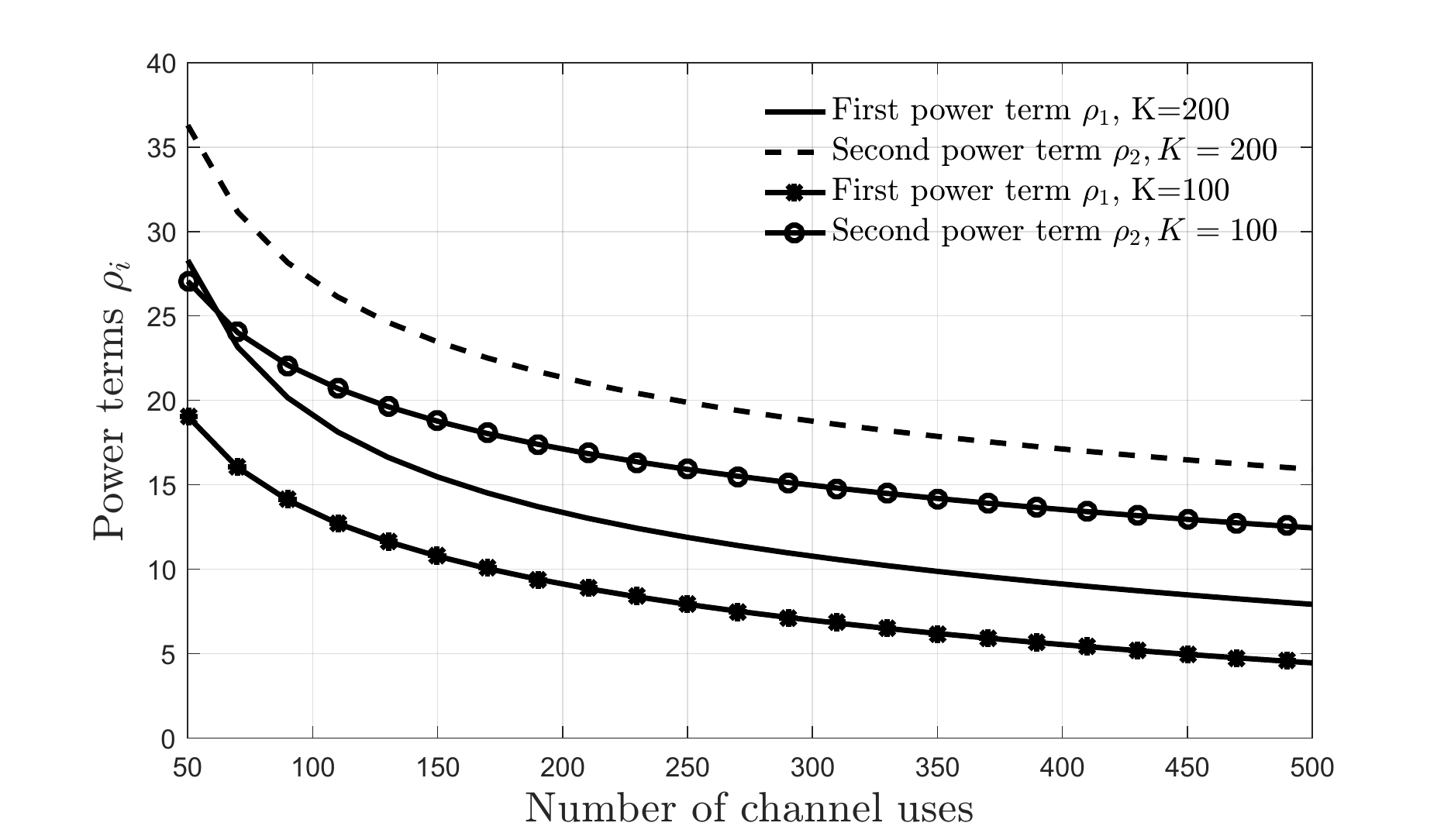}
		\caption{Power terms as a function of the number of channel uses for different number of information bits.}
		\label{fig:power_terms_vs_n_and_k}
		%\vspace*{-2mm}
	\end{figure}

	Fig. \ref{fig:power_terms_vs_n_and_k} illustrates the behaviour of the power terms as a function of the number of channel uses when $M=2$. Here we set the number of information bits $K$ to 100 and 200. First we observe that both power terms decrease as we increase the number of channel uses. Secondly we notice that when we increase the coding rate, we have to transmit with higher power in each ARQ round.
	
\begin{figure}[!htb] % [!t] or [!b] or [!h] % force fitting, force top, force bottom, force text fitting
		\vspace*{-3.5mm}
		\centering
		\includegraphics[trim=0.8cm 0.1cm 1.3cm 0.7cm, clip=true,width=.8\linewidth]{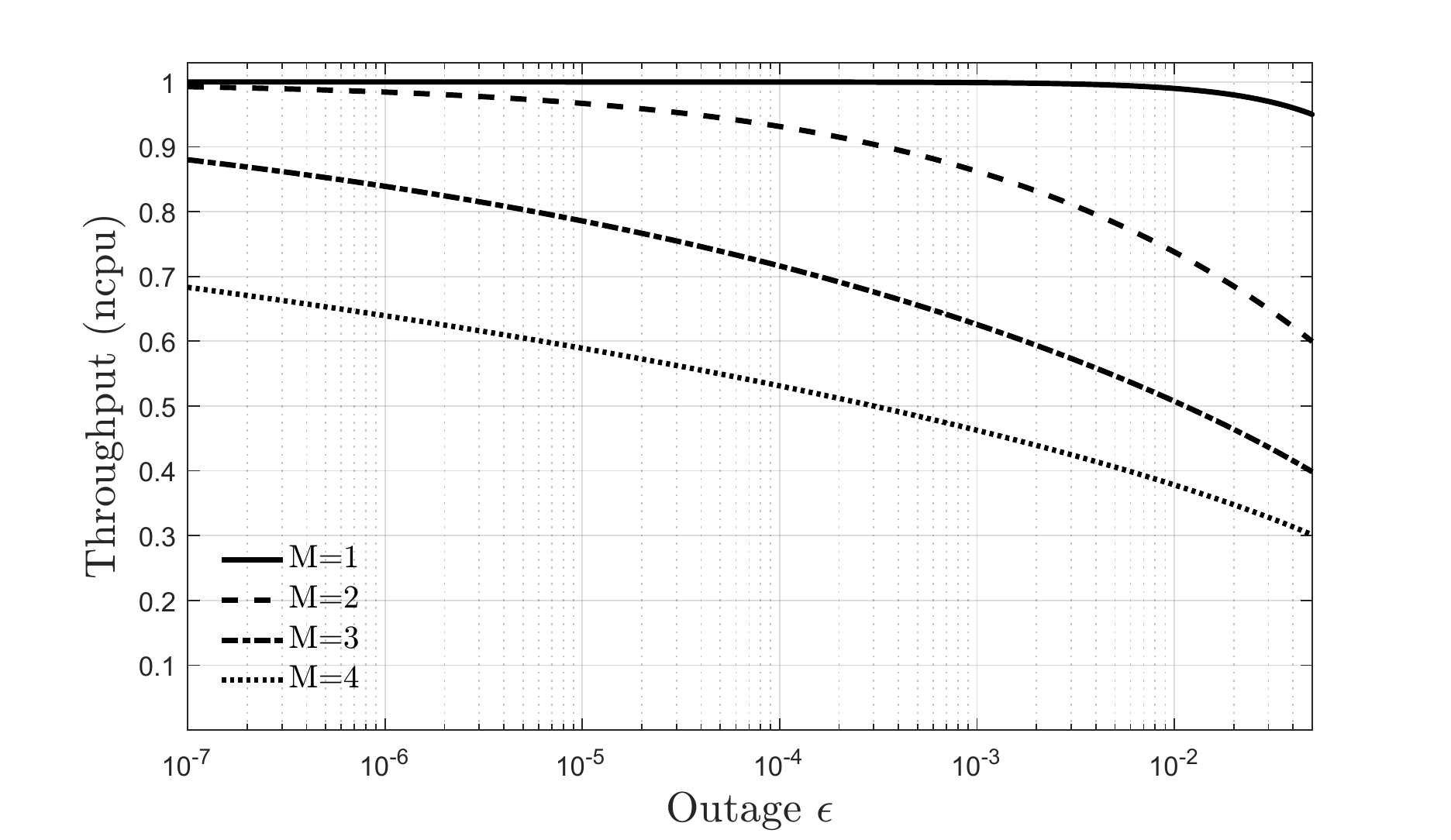}
		\caption{Throughput as a function of outage in the ultra reliable region.}
		\label{fig:Througthput_vs_outage}
		%\vspace*{-2mm}
	\end{figure}

	The corresponding outage terms for the power allocation scheme are computed by substituting the power terms in \eqref{eq:14}, and they are given by $\epsilon_1= \sqrt[3]{\tfrac{\epsilon}{2}}$ and $\epsilon_2= \epsilon\sqrt[3]{\tfrac{2}{\epsilon}}$. To calculate the throughput we substitute these outage terms and $M=2$ in \eqref{eq:31}. The behavior of the throughput as a function of the target outage probability for different number of retransmissions is illustrated in Fig. \ref{fig:Througthput_vs_outage}. Here, we set $R=1$ ncpu and assume no feedback delay, thus $D=0$. Notice that there is a loss in the throughput due to the ARQ protocol, but the proposed power allocation scheme with $M=2$ helps in mitigating this loss for very low outage values. Furthermore, the results show that as we increase the number of retransmissions, the throughput starts to decrease. We observe that when $M=3$ the throughput decreases by 11\% with respect to the open loop setup. When $M=4 $  the decrease is 21\%. 
	
\begin{figure}[!htb] % [!t] or [!b] or [!h] % force fitting, force top, force bottom, force text fitting
	%	\vspace*{-3mm}
		\centering
		\includegraphics[trim=0.8cm 0.1cm 1.3cm 0.7cm, clip=true,width=.8\linewidth]{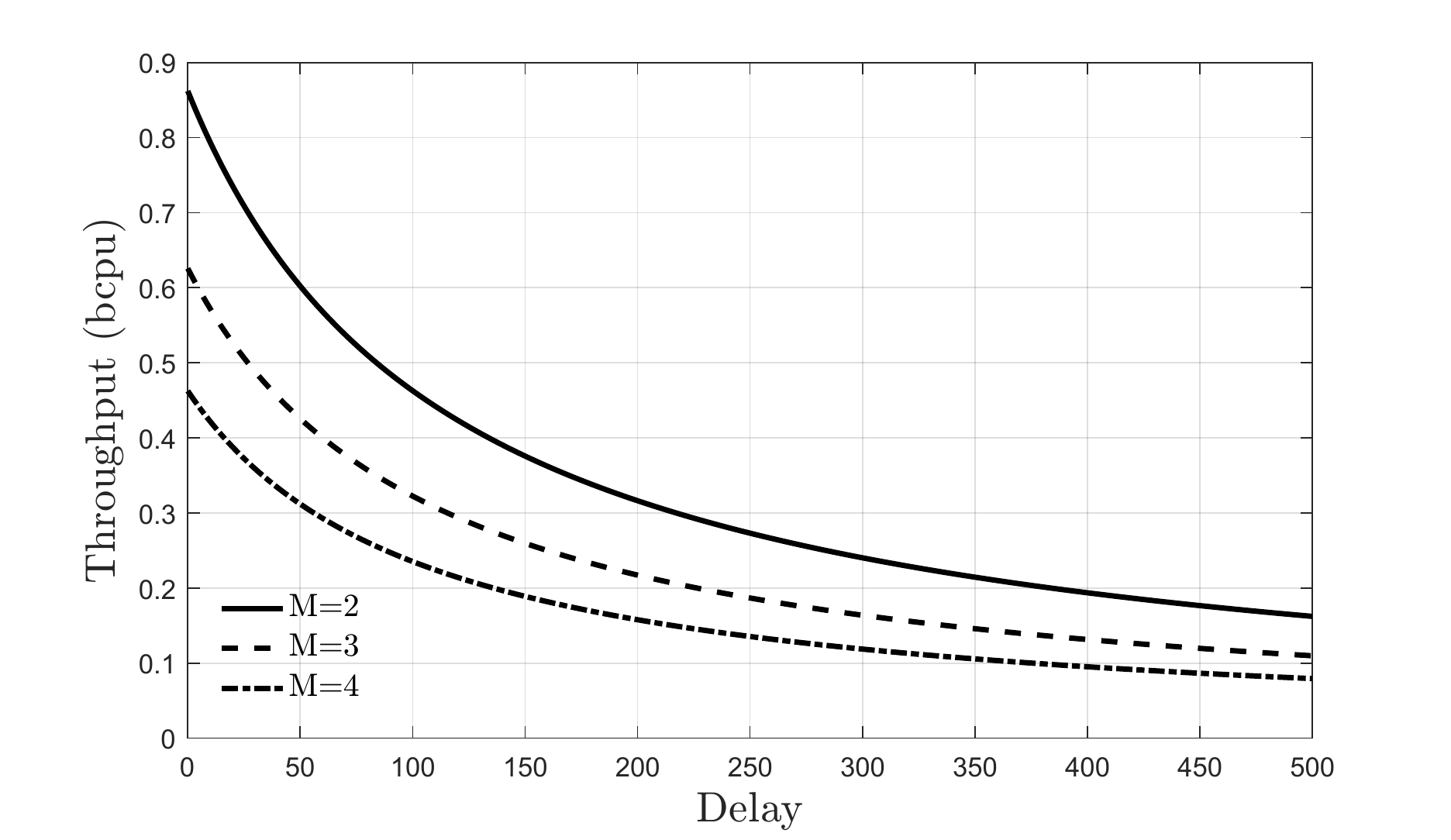}
		\caption{Throughput as a function of delay in the ultra reliable region.}
		\label{fig:Througthput_vs_delay}
		
	\end{figure}

	The behavior of the throughput as a function of delay is investigated in Fig. \ref{fig:Througthput_vs_delay}. The results are shown for coding rate $R=1$ ncpu and target outage probability $\epsilon=10^{-3}$. From Fig. \ref{fig:Througthput_vs_delay} we can see that the retransmission and feedback delays have a great impact in the throughput for a fixed $\epsilon$ despite the number of retransmissions. However, in ultra reliable systems this loss can be tolerated, especially when considering the larger gains attained through the proposed power allocation scheme, as observed in Fig.\ref{fig:Power_terms_vs_outage}. 

	\section{Conclusions and future work}
	\label{5}
	In this paper, we have developed an optimal power allocation scheme for the type-I ARQ protocol that allows to operate in the ultra reliable region while spending minimal power. We have shown that our proposed power allocation scheme can maximize the overall system throughput as well, with or without the presence of feedback delay. As future work, we intend to analyze the performance of other protocols such as hybrid-ARQ with power allocation schemes in the finite block-length regime and also analyze how the optimal power allocation scheme changes when consider the case with feedback delay constraints.
	\section*{Acknowledgments}
	This work has been partially supported by Finnish Funding Agency for Technology and Innovation (Tekes), Huawei Technologies, Nokia and Anite Telecoms, and Academy of Finland.
	\vspace*{-3mm}
	%	
	% Bibliographic styles 
	\bibliographystyle{IEEEtran}
	\bibliography{IEEEabrv,Papers}

\begin{thebibliography}{10}
\providecommand{\url}[1]{#1}
\csname url@samestyle\endcsname
\providecommand{\newblock}{\relax}
\providecommand{\bibinfo}[2]{#2}
\providecommand{\BIBentrySTDinterwordspacing}{\spaceskip=0pt\relax}
\providecommand{\BIBentryALTinterwordstretchfactor}{4}
\providecommand{\BIBentryALTinterwordspacing}{\spaceskip=\fontdimen2\font plus
\BIBentryALTinterwordstretchfactor\fontdimen3\font minus
  \fontdimen4\font\relax}
\providecommand{\BIBforeignlanguage}[2]{{%
\expandafter\ifx\csname l@#1\endcsname\relax
\typeout{** WARNING: IEEEtran.bst: No hyphenation pattern has been}%
\typeout{** loaded for the language `#1'. Using the pattern for}%
\typeout{** the default language instead.}%
\else
\language=\csname l@#1\endcsname
\fi
#2}}
\providecommand{\BIBdecl}{\relax}
\BIBdecl

\bibitem{Popovski2014}
P.~Popovski, ``Ultra-reliable communication in 5{G} wireless systems,'' in
  \emph{International Conference on 5{G} for Ubiquitous Connectivity}, Nov
  2014.

\bibitem{Networks}
{Nokia Networks }, ``{LTE-M} - optimizing {LTE} for the {I}nternet of
  {T}hings,'' \emph{Nokia networks white paper}, 2015.

\bibitem{Palattella}
M.~R. Palattella, M.~Dohler, A.~Grieco, G.~Rizzo, J.~Torsner, T.~Engel, and
  L.~Ladid, ``Internet of things in the 5{G} era: Enablers, architecture, and
  business models,'' \emph{IEEE Journal on Selected Areas in Communications},
  March 2016.

\bibitem{Andreev}
S.~Andreev, O.~Galinina, A.~Pyattaev, M.~Gerasimenko, T.~Tirronen, J.~Torsner,
  J.~Sachs, M.~Dohler, and Y.~Koucheryavy, ``Understanding the {I}o{T}
  connectivity landscape: a contemporary {M2M} radio technology roadmap,''
  \emph{IEEE Communications Magazine}, vol.~53, no.~9, pp. 32--40, September
  2015.

\bibitem{Nokia}
{Nokia Networks}, ``5{G} for mission critical communication: Achieve
  ultra-reliability and virtual zero latency,'' \emph{Nokia networks white
  paper}, 2016.

\bibitem{Goldsmith}
A.~Goldsmith, \emph{{Wireless Communications}}.\hskip 1em plus 0.5em minus
  0.4em\relax Cambridge, UK: Cambridge University Press, 2005.

\bibitem{Larsson}
P.~Larsson, L.~K. Rasmussen, and M.~Skoglund, ``Analysis of rate optimized
  throughput for large-scale {MIMO-(H)ARQ} schemes,'' in \emph{IEEE Globecom},
  Dec 2014.

\bibitem{Chaitanya2016}
T.~V.~K. Chaitanya and T.~Le-Ngoc, ``{Energy-efficient adaptive power
  allocation for incremental MIMO systems},'' \emph{IEEE Transactions on
  Vehicular Technology}, 2016.

\bibitem{Tumula2}
T.~V.~K. Chaitanya and E.~G. Larsson, ``Optimal power allocation for hybrid
  {ARQ} with chase combining in i.i.d. {R}ayleigh fading channels,'' \emph{IEEE
  Transactions on Communications}, May 2013.

\bibitem{Durisi_1}
G.~Durisi, T.~Koch, and P.~Popovski, ``Toward massive, ultra-reliable, and
  low-latency wireless communication with short packets,'' \emph{Proceedings of
  the IEEE}, Sept 2016.

\bibitem{Kim2013}
S.~H. Kim, D.~K. Sung, and T.~Le-Ngoc, ``{Performance analysis of incremental
  redundancy type hybrid ARQ for finite-length packets in AWGN channel},'' in
  \emph{IEEE GLOBECOM}, 2013, pp. 2063--2068.

\bibitem{Devassy2014}
R.~Devassy, G.~Durisi, P.~Popovski, and E.~G. Strom, ``{Finite-blocklength
  analysis of the ARQ-protocol throughput over the Gaussian collision
  channel},'' \emph{IEEE ISCCSP}, pp. 173--177, 2014.

\bibitem{Makki2015}
B.~Makki, T.~Svensson, and M.~Zorzi, ``{Green communication via Type-I ARQ:
  Finite block-length analysis},'' in \emph{IEEE GLOBECOM}, 2014.

\bibitem{Polyanskiy2010}
Y.~Polyanskiy, H.~V. Poor, and S.~Verdu, ``Channel coding rate in the finite
  blocklength regime,'' \emph{IEEE Transactions on Information Theory},
  vol.~56, no.~5, May 2010.

\bibitem{Yanga}
W.~Yang, G.~Durisi, T.~Koch, and Y.~Polyanskiy, ``Quasi-static multiple-antenna
  fading channels at finite blocklength,'' \emph{IEEE Transactions on
  Information Theory}, vol.~60, no.~7, pp. 4232--4265, July 2014.

\bibitem{Yangk}
------, ``Block-fading channels at finite blocklength,'' in \emph{Wireless
  Communication Systems (ISWCS 2013), Proceedings of the Tenth International
  Symposium on}, Aug 2013.

\bibitem{Makki2014}
B.~Makki, T.~Svensson, and M.~Zorzi, ``{Finite block-length analysis of the
  incremental redundancy HARQ},'' \emph{IEEE Wireless Communications Letters},
  vol.~3, no.~5, pp. 529--532, 2014.

\bibitem{5754756}
T.~V.~K. Chaitanya and E.~G. Larsson, ``Outage-optimal power allocation for
  hybrid {ARQ} with incremental redundancy,'' \emph{IEEE Transactions on
  Wireless Communications}, vol.~10, no.~7, July 2011.

\bibitem{Boyd-Vandenberghe-04}
S.~Boyd and L.~Vandenberghe, \emph{Convex Optimization}.\hskip 1em plus 0.5em
  minus 0.4em\relax Cambridge, UK: Cambridge University Press, 2004.

\end{thebibliography}
	
	% - - - - - - - - - - - - - - - - - - - - - - - - - - - - - - - - - - - - - - - - - - - - - - 
\end{document}